\documentclass[
  journal=pasa,
  manuscript=research-paper 
]{cup-journal}

\usepackage{microtype,siunitx,booktabs}
\usepackage{amsmath, amsthm, amssymb, commath, graphicx, bbold, endnotes, graphicx, subfigure, multirow, comment}

\title{The \textsc{fhd} polarized imaging pipeline: A new approach to widefield interferometric polarimetry}

\author{Ruby L.\ Byrne}
\affiliation{Astronomy Department, California Institute of Technology, 1200 E California Blvd, Pasadena, CA, 91125, USA}
\email[Ruby L.\ Byrne]{rbyrne@caltech.edu}

\author{Miguel F.~Morales}
\affiliation{Physics Department, University of Washington, 3910 15th Ave NE, Seattle, WA, 98195, USA}

\author{Bryna Hazelton}
\affiliation{Physics Department, University of Washington, 3910 15th Ave NE, Seattle, WA, 98195, USA}
\alsoaffiliation{eScience Institute, University of Washington, 3910 15th Ave NE, Seattle, WA, 98195, USA}

\author{Ian Sullivan}
\affiliation{Astronomy Department, University of Washington, 3910 15th Ave NE, Seattle, WA, 98195, USA}

\author{Nichole Barry}
\affiliation{International Centre for Radio Astronomy Research, Curtin University, Perth, WA 6845, Australia}
\alsoaffiliation{Australian Research Council Centre of Excellence for All Sky Astrophysics in 3 Dimensions (ASTRO 3D), Australia}

\received {25 Jan 2022}
\revised  {30 Mar 2022}
\accepted {13 Apr 2022}
\published{13 May 2022}


\begin{document}

\newcommand\fitparams{\boldsymbol{\xi}}
\newcommand\fitparam{\xi}
\newcommand\freq{f}
\newcommand\delay{\eta}
\newcommand\visfunc{\boldsymbol{\zeta}}
\newcommand\visfuncscalar{\zeta}
\newcommand\freqparams{\boldsymbol{\gamma}}
\newcommand\freqparam{\gamma}
\newcommand\thermalcov{\boldsymbol{\mathsf{C}}_\text{T}}
\newcommand\modelcov{\boldsymbol{\mathsf{C}}_\text{M}}
\newcommand\modelcovscalar{C_\text{M}}
\newcommand\data{\boldsymbol{v}}
\newcommand\datascalar{v}
\newcommand\fisherinfo{\boldsymbol{\mathsf{I}}}
\newcommand\fisherinfoscalar{I}
\newcommand\gains{\boldsymbol{g}}
\newcommand\gain{g}
\newcommand\gainsmat{\boldsymbol{\mathsf{G}}}
\newcommand\fitparamsu{\boldsymbol{u}}
\newcommand\fitparamu{u}
\newcommand\modelvals{\boldsymbol{m}}
\newcommand\modelval{m}
\newcommand\thermalvar{\sigma_\text{T}^2}
\newcommand\thermalvarvector{\boldsymbol{\sigma}_\text{T}^2}
\newcommand\modelvar{\sigma_\text{M}^2}
\newcommand\modelvarvector{\boldsymbol{\sigma}_\text{M}^2}
\newcommand\matrixa{\boldsymbol{\mathsf{A}}}
\newcommand\ascalar{A}
\newcommand\uvcoord{\boldsymbol{u}} 
\newcommand\beamscalar{B}
\newcommand\beamvec{\boldsymbol{B}}
\newcommand\beammat{\boldsymbol{\mathsf{B}}}
\newcommand\jonesmat{\boldsymbol{\mathsf{J}}}
\newcommand\jonesscalar{J}
\newcommand\coherencyscalar{S}
\newcommand\coherency{\boldsymbol{S}}
\newcommand\electricfield{E}
\newcommand\vismat{\boldsymbol{\mathsf{V}}}
\newcommand\negloglikelihood{L}
\newcommand\racoord{\text{R}}
\newcommand\deccoord{\text{D}}
\newcommand\muellermat{\boldsymbol{\mathsf{M}}}
\newcommand\unitvec{\boldsymbol{e}}
\newcommand\basistransform{\boldsymbol{\mathsf{K}}}
\newcommand\muellerbasistransform{\boldsymbol{\mathsf{L}}}
\newcommand\rotangle{\beta}
\newcommand\rotmeas{R}
\newcommand\wavelength{\lambda}
\newcommand\skypos{\boldsymbol{\theta}}
\newcommand\holographplane{\boldsymbol{P}_\text{holo}}
\newcommand\uniformplane{\boldsymbol{P}_\text{uni}}
\newcommand\weights{\boldsymbol{W}}
\newcommand\numvis{\boldsymbol{N}}
\newcommand\fillingfactor{C}
\newcommand\obsindex{n}
\newcommand\radweight{W}
\newcommand\gainamp{A}
\newcommand\note[1]{\textcolor{red}{#1}}
\newcommand\groundcoord{\boldsymbol{r}}
\newcommand\antres{F}
\newcommand\antresponse{\epsilon}
\newcommand\antresmat{\boldsymbol{\mathsf{F}}}
\newcommand\unnormgains{\boldsymbol{h}}
\newcommand\unnormgain{h}
\newcommand\modelcorr{\boldsymbol{\mathsf{C}}_\text{R}}
\newcommand\modelcorrscalar{C_\text{R}}
\newcommand\efield{E}
\newcommand\transferfunc{T}
\newcommand\transferfuncmat{\boldsymbol{\mathsf{T}}}
\newcommand\ewinstpol{\text{p}}
\newcommand\nsinstpol{\text{q}}
\newcommand\parallacticangle{\phi}
\newcommand\raval{\alpha}
\newcommand\decval{\delta}
\newcommand\noise{n}
\newcommand\ft{\mathcal{FT}}
\newcommand\re{\operatorname{Re}}
\newcommand\im{\operatorname{Im}}
\newcommand\overallphase{\Delta}
\newcommand\argument{\operatorname{Arg}}
\newcommand\overallamp{A}
\newcommand\phasegradx{\Delta_x}
\newcommand\phasegrady{\Delta_y}
\newcommand\relcalgains{\boldsymbol{h}}
\newcommand\relcalgain{h}
\newcommand\crosspolphase{\Delta}

\begin{abstract}
    We describe a new polarized imaging pipeline implemented in the \textsc{fhd} software package. The pipeline is based on the optimal mapmaking imaging approach and performs horizon-to-horizon image reconstruction in all polarization modes. We discuss the formalism behind the pipeline's polarized analysis, describing equivalent representations of the polarized beam response, or Jones matrix. We show that, for arrays where antennas have uniform polarization alignments, defining a non-orthogonal instrumental polarization basis enables accurate and efficient image reconstruction. Finally, we present a new calibration approach that leverages widefield effects to perform fully-polarized calibration. This analysis pipeline underlies the analysis of Murchison Widefield Array (MWA) data in \citealt{Byrne2022}.
\end{abstract}

\section{Introduction}

The field of low-frequency radio astronomy has expanded in recent years with the development of radio arrays such as the Low-Frequency Array (LOFAR; \citealt{VanHaarlem2013}), the Murchison Widefield Array (MWA; \citealt{Tingay2013}), the Giant Metrewave Radio Telescope (GMRT; \citealt{Swarup1990}), the Donald C. Backer Precision Array for Probing the Epoch of Reionization (PAPER; \citealt{Parsons2010}), the Hydrogen Epoch of Reionization Array (HERA; \citealt{DeBoer2014}), the Owens Valley Radio Observatory Long Wavelength Array (OVRO-LWA; \citealt{Eastwood2018}; \citealt{Anderson2019}), the forthcoming Square Kilometre Array (SKA; \citealt{Mellema2013}), and others. These powerful instruments are expanding radio astronomy observations to lower frequencies with enhanced sensitivity and improved resolution. They are inherently widefield instruments with sensitivity across large swaths of the sky. This widefield imaging regime has necessitated the development of novel interferometric data analysis techniques to confront the challenges of horizon-to-horizon image reconstruction (\citealt{Bhatnagar2008, Cornwell2008, Morales2009, Tasse2013, Offringa2014}; and others). A particular challenge is the widefield reconstruction of fully polarized images. Here we describe a new approach to widefield polarized imaging and calibration, implemented as a capability of the Fast Holographic Deconvolution (\textsc{fhd}) software pipeline\footnote{\texttt{https://github.com/EoRImaging/FHD}} \citep{Sullivan2012, Barry2019a}.

\textsc{fhd} is a versatile interferometric data analysis package written in the IDL programming language that performs calibration, imaging, data simulation, and compact source deconvolution. The polarized imaging pipeline described in this paper was applied to MWA data to map diffuse emission across much of the Southern Hemisphere sky \citep{Byrne2022}. For a discussion of the computational requirements of that analysis using Amazon Web Services (AWS) cloud-based instances, see \citealt{Byrne2021a}. This paper serves as a companion to \citealt{Sullivan2012} and \citealt{Barry2019a}; together, these three papers describe various aspects of the \textsc{fhd} data processing pipeline.

\textsc{fhd}'s analysis is an implementation of the optimal mapmaking imaging approach developed by \citealt{Bhatnagar2008} and \citealt{Morales2009} and based upon the formalism developed in \citealt{Tegmark1997}. Under optimal mapmaking, which also goes by the name ``A-projection,'' visibilities are gridded to a \textit{uv} plane with a kernel defined by the instrumental response beam. This approach accounts for widefield effects and accurately reconstructs images across the entire sky. Here we describe a fully-polarized extension to this imaging approach that produces horizon-to-horizon images in all four Stokes polarization parameters.

This implementation performs image reconstruction in the ``instrumental'' polarization basis, described in detail in \S\ref{s:instr_basis}. The images can then be transformed into the usual Stokes polarization modes with no loss of information. This approach has significant computational benefits and is applicable to any array in which all antennas have identical polarization alignments. An instrument with antennas that are rotated with respect to one another cannot exploit the instrumental polarization basis for efficient imaging.

The \textsc{fhd} polarization pipeline builds upon the work of other widefield polarized imagers. Notably, it is similar to the A-projection LOFAR analysis pipeline described in \citealt{Tasse2013}. It also shares features with \textsc{wsclean} \citep{Offringa2014} and the MWA's \textsc{rts} analysis pipeline \citep{Mitchell2012}. Although the implementations differ in the specifics of their image estimation, all either implicitly or explicitly reconstruct images in the instrumental basis before correcting for widefield projection effects. 

This paper is intended to illuminate the details of \textsc{fhd}'s implementation while providing a comprehensive exploration of widefield polarimetric imaging and the instrumental basis. It offers explanation of the analysis underlying \citealt{Byrne2022} and future \textsc{fhd}-based polarization studies and situates \textsc{fhd} within the broader field of widefield interferometric polarimetry.

In the next section we define polarimetric terms and relationships that are used throughout the paper. In \S\ref{s:instr_basis} we discuss the instrumental basis, and in \S\ref{s:polarized_imaging} we describe \textsc{fhd}'s polarized image estimation. \S\ref{s:polarized_calibration} presents \textsc{fhd}'s polarized calibration implementation.

\section{Polarization Formalism}
\label{s:polarization_formalism}

In this section we present an overview of some of the basic terms and relationships used throughout this paper: the coherency, the Stokes parameters, the Jones matrix, and the Mueller matrix.

\subsection{The Coherency}
\label{s:coherency}

We describe the proprieties of the electric field signal from the sky with a coherency vector, given by
\begin{equation}
    \coherency(\skypos) = \begin{bmatrix}
    \langle |\electricfield_\racoord(\skypos)|^2 \rangle \\ \langle |\electricfield_\deccoord(\skypos)|^2 \rangle \\
    \langle \electricfield_\racoord(\skypos) \electricfield_\deccoord^*(\skypos) \rangle \\
    \langle \electricfield_\racoord^*(\skypos) \electricfield_\deccoord(\skypos) \rangle
    \end{bmatrix}
\label{eq:coherency_def}
\end{equation}
\citep{Hamaker1996a}. Here $\electricfield_\racoord(\skypos)$ and $\electricfield_\deccoord(\skypos)$ are the components of the electric field in two orthogonal directions. In our convention, we define these directions to align with the RA and Dec.\ directions on the sky, respectively. $\skypos$ is a two-element vector defining the position on the sky. The angled brackets $\langle \rangle$ denote the time average and the asterisk $^*$ represents the complex conjugate.

The coherency can equivalently be described as a $2\times2$ matrix \citep{Hamaker1996a, Smirnov2011}. Likewise, the vector ordering and orthogonal basis are arbitrary. Equation \ref{eq:coherency_def} presents the convention used in the \textsc{fhd} analysis.

\subsection{The Stokes Parameters}
\label{s:stokes}

Polarized emission is often described with respect to the Stokes parameters I, Q, U, and V \citep{Stokes1851}. Stokes I corresponds to the total intensity, Stokes Q and U correspond to linear polarization, and Stokes V corresponds to circular polarization. The Stokes parameters are related to the coherency vector $\coherency(\skypos)$ via the relationship
\begin{equation}
    \begin{bmatrix}
    I(\skypos) \\
    Q(\skypos) \\
    U(\skypos) \\
    V(\skypos)
    \end{bmatrix} = \begin{bmatrix}
	1 & 1 & 0 & 0 \\
	1 & -1 & 0 & 0 \\
	0 & 0 & 1 & 1 \\
	0 & 0 & i & -i \\
	\end{bmatrix} \coherency(\skypos).
\label{eq:stokes_def}
\end{equation}

Note that, because Equation \ref{eq:coherency_def} defines the coherency vector with respect to the RA/Dec.\ coordinate system, the Stokes parameters are also referenced to that coordinate system. We emphasize that the Stokes parameters are basis-dependent, and it is critical that any polarized measurements specify the basis used \citep{Hamaker1996b}. Under the RA/Dec.\ coordinate system, the Stokes parameters are undefined at the North and South Celestial Poles. Consequently, analyses of fields at or near the poles may benefit from choosing a different orthogonal basis. See \citealt{Ludwig1973} for examples of alternative polarization bases that may be used to define the Stokes parameters.

\subsection{The Jones Matrix}

The Jones matrix is a $2\times2$ complex matrix that represents the polarized antenna response to the sky \citep{Jones1941}. The matrix transforms between the true electric field on the sky and the electric field measured by the instrument \citep{Hamaker1996a, Sault1996, Hamaker1996b, Hamaker2000, Hamaker2006, Ord2010, Smirnov2011}:
\begin{equation}
    \begin{bmatrix}
    \antresponse_{j \ewinstpol}(\skypos) \\
    \antresponse_{j \nsinstpol}(\skypos)
    \end{bmatrix} = \jonesmat^\text{ZA}_j(\skypos) \begin{bmatrix}
    \electricfield_\text{Z}(\skypos) \\
    \electricfield_\text{A}(\skypos)
    \end{bmatrix}.
\label{eq:jones_mat_za}
\end{equation}
Here $\electricfield_\text{Z}(\skypos)$ and $\electricfield_\text{A}(\skypos)$ are components of the electric field aligned with the zenith angle and azimuth directions, respectively. 
$\antresponse_{j \ewinstpol}(\skypos)$ and $\antresponse_{j \nsinstpol}(\skypos)$ represent the contribution of emission from sky direction $\skypos$ to the measurements made by antenna $j$, and the subscripts $\ewinstpol$ and $\nsinstpol$ correspond to the two instrumental polarization modes. For example, for an antenna with dipole elements aligned with the cardinal directions, $\ewinstpol$ could refer to the east-west aligned dipole and $\nsinstpol$ could refer to the north-south aligned dipole. $\jonesmat^\text{ZA}_j(\skypos)$ is the Jones matrix for antenna $j$. The superscript ZA indicates that it corresponds to the zenith angle/azimuth basis on the sky. This is the usual basis convention for reporting the Jones matrix.

This formalism is fully general in that it assumes dual-polarization antenna but makes no further assumptions about the antenna type. Each antenna could consist of dish with a dual-polarization feed (e.g.\ GMRT or HERA), a simple crossed-dipole element (as with PAPER or the OVRO-LWA), or a beamformed array of dipole elements (such as LOFAR's stations or the MWA's tiles). The two antenna polarizations could be orthogonal but need not be. They could measure linearly or circularly polarized modes. Note that here we have defined a per-antenna Jones matrix that depends on the antenna index $j$: antennas need not have homogeneous polarized responses. The Jones matrix is direction-dependent, and we require a $2\times2$ Jones matrix for each location on the sky $\skypos$. It is also complex-valued: the direction-dependent complex phase of the Jones matrix elements captures timing delays in the instrumental response to an incoming wavefront.

\begin{figure}
    \centering
    \includegraphics[width=\columnwidth]{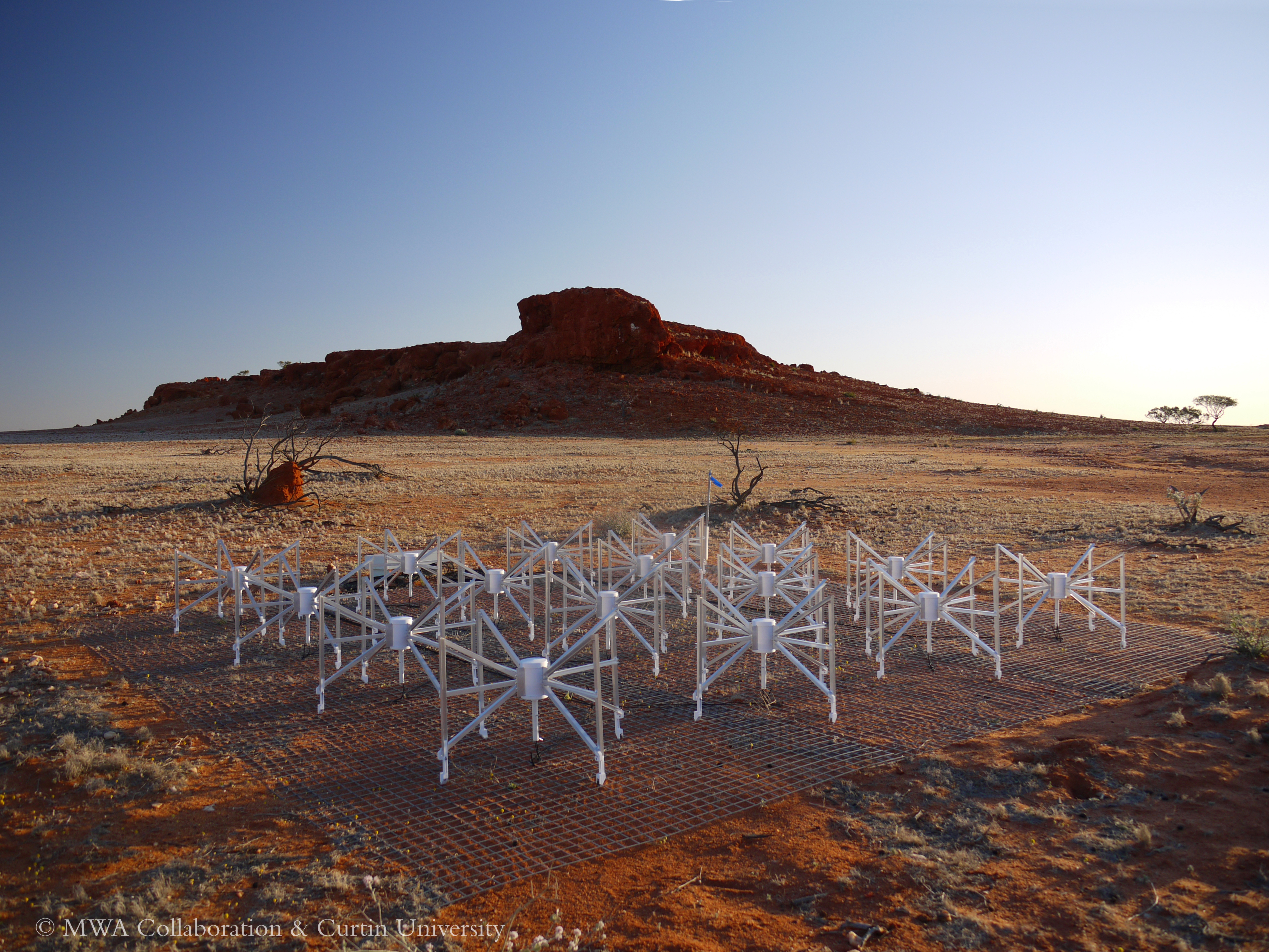}
    \caption{A photo of an MWA tile. Each tile consists of 16 dual-polarization beamformed elements, and the full array consists of 256 such tiles \citep{Wayth2018}. The tile's response to incident radiation is estimated by a beam model. Three equivalent representations of the beam model are shown in each Figure \ref{fig:jones_za}, Figure \ref{fig:jones_ra}, and Figures \ref{fig:beam_amp} and \ref{fig:instr_basis}. Photo credit: Natasha Hurley-Walker, the MWA Collaboration, and Curtin University.}
    \label{fig:mwa_tile}
\end{figure}

\begin{figure*}
    \centering
    \includegraphics[width=0.7\columnwidth]{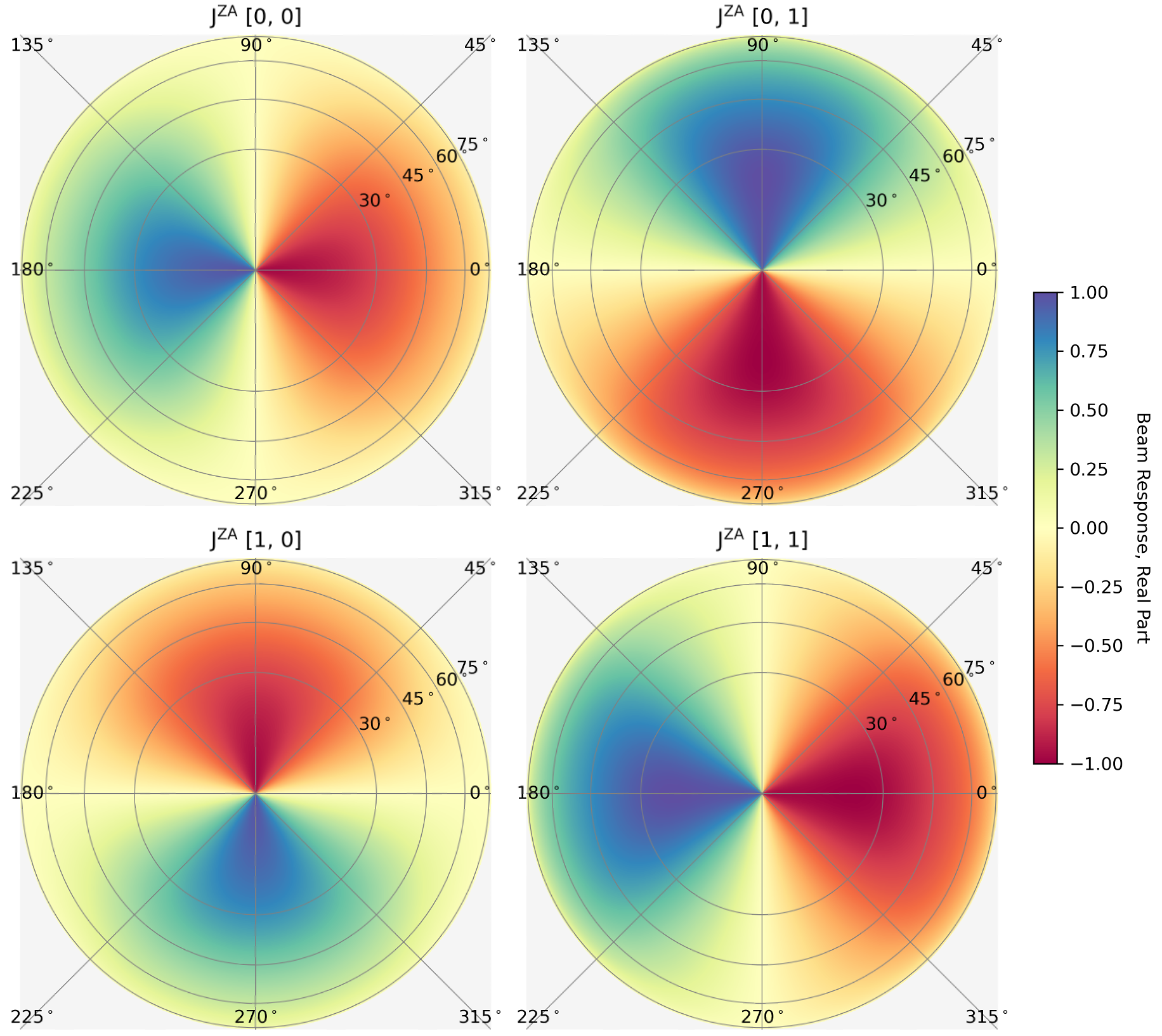}
    \caption{The elements of the Jones matrix for a zenith-pointed MWA tile (pictured in Figure 
    \ref{fig:mwa_tile}) at 167-198 MHz, as modeled by \citealt{Sutinjo2015}. The Jones matrix defines and instrument's polarized response. It is a $2\times2$ complex matrix defined at each point on the sky; here we plot the real part only. The top row depicts the response of the east-west aligned, or $\ewinstpol$, tile polarization while the bottom row depicts the response of the north-south aligned, or $\nsinstpol$, tile polarization. Here the Jones matrix is normalized such that the peak response amplitude of each tile polarization is one. This Jones matrix is defined with respect to the zenith angle/azimuth basis (see Equation \ref{eq:jones_mat_za}). The left and right columns corresponds to the tile's response to electric field emission polarized in the zenith angle and azimuth directions, respectively. The Jones matrix elements exhibit a discontinuity at zenith as a result of the pole of the coordinate system.}
    \label{fig:jones_za}
\end{figure*}

In Figure \ref{fig:jones_za} we plot the four elements of the Jones matrix of an MWA tile, as modeled by \citet{Sutinjo2015}, as a function of position on the sky. Each MWA tile, pictured in Figure \ref{fig:mwa_tile}, consists of 16 dual-polarization beamformed elements. The Jones matrix model in Figure \ref{fig:jones_za} corresponds to the full tile response in its zenith-pointed mode, averaged across a frequency range of 167-198 MHz. This Jones matrix defines the instrumental response for \textsc{fhd} analyses including the polarized mapping in \citet{Byrne2022} and the EoR analyses in \citet{Barry2019b} and \citet{Li2019}. While the Jones matrix is complex-valued, the complex phase is near-zero, and for simplicity we plot the real part only. Note that elements of the Jones matrix exhibit a discontinuity at zenith. This is a feature of the basis in which it is defined and not a characteristic of the physical antenna response.

Equation \ref{eq:jones_mat_za} and Figure \ref{fig:jones_za} present the Jones matrix in the zenith angle/azimuth coordinate system, but in Equation \ref{eq:coherency_def} we define the coherency with respect to the RA/Dec.\ basis. Our analysis is therefore simplified if we define a Jones matrix $\jonesmat^{\racoord \deccoord}_j(\skypos)$ with respect to RA/Dec. The transformation between bases involves rotating by the direction-dependent parallactic angle $\parallacticangle(\skypos)$:
\begin{equation}
    \begin{bmatrix}
    \unitvec_Z(\skypos) \\
    \unitvec_A(\skypos)
    \end{bmatrix} = \begin{bmatrix} 
	\sin [\parallacticangle(\skypos)] & -\cos [\parallacticangle(\skypos)] \\
	-\cos [\parallacticangle(\skypos)] & -\sin [\parallacticangle(\skypos)]
	\end{bmatrix} 
	\begin{bmatrix}
    \unitvec_\racoord(\skypos) \\
    \unitvec_\deccoord(\skypos)
    \end{bmatrix}.
\label{eq:parallactic_angle_rot}
\end{equation}
Here $\unitvec_\racoord(\skypos)$, $\unitvec_\deccoord(\skypos)$, $\unitvec_Z(\skypos)$, and $\unitvec_A(\skypos)$ are unit vectors in the RA, Dec., zenith angle, and azimuth directions, respectively. The parallactic angle is given by
\begin{equation}
    \parallacticangle(\skypos) = \tan^{-1} \left( \frac{-\sin \raval}{\cos \decval \tan \decval_\text{zen} - \sin \decval \cos \raval} \right),
\label{eq:par_angle}
\end{equation}
where $\raval$ and $\decval$ are the RA and Dec.\ coordinates of $\skypos$, respectively, and $\decval_\text{zen}$ is the declination at zenith. For a full derivation of this expression see \citealt{Byrne_thesis}, Appendix D.

Here Equation \ref{eq:par_angle} assumes that the Earth's axis of rotation aligns with the poles of the RA/Dec.\ coordinate system. Under this assumption the parallactic angle is time-independent: for an array at a given declination, the transformation between the zenith angle/azimuth and RA/Dec.\ polarization bases does not depend on time. In practice, the Earth's precession and nutation introduce deviations in the alignment of the rotational axis. While \textsc{fhd} fully accounts for precession and nutation when calculating source positions on the sky, where small positional errors can substantially degrade analysis accuracy, it does not account for them when transforming between polarization bases, instead using the parallactic angle calculation presented in Equation \ref{eq:par_angle}. This is a good approximation as the errors from the neglected precession and nutation effects produce only small errors in the reconstructed polarization angle. Future extensions to \textsc{fhd}'s polarized imaging pipeline could add these higher-order effects to the calculation of the polarization basis transformation. This would introduce time dependence to the parallactic angle calculation in Equation \ref{eq:par_angle} and by extension the basis transformation in Equation \ref{eq:parallactic_angle_rot}.

Applying the transformation in Equation \ref{eq:parallactic_angle_rot} to the Jones matrix, we get that
\begin{equation}
    \jonesmat^{\racoord \deccoord}_j(\skypos) = \jonesmat^\text{ZA}_j(\skypos) \begin{bmatrix} 
	\sin [\parallacticangle(\skypos)] & -\cos [\parallacticangle(\skypos)] \\
	-\cos [\parallacticangle(\skypos)] & -\sin [\parallacticangle(\skypos)]
	\end{bmatrix}
\label{eq:jones_rot}
\end{equation}
where
\begin{equation}
    \begin{bmatrix}
    \antresponse_{j \ewinstpol}(\skypos) \\
    \antresponse_{j \nsinstpol}(\skypos)
    \end{bmatrix} = \jonesmat^{\racoord \deccoord}_j(\skypos) \begin{bmatrix}
    \electricfield_\racoord(\skypos) \\
    \electricfield_\deccoord(\skypos)
    \end{bmatrix}.
\label{eq:jones_mat_radec}
\end{equation}
$\jonesmat^{\racoord \deccoord}_j(\skypos)$, the Jones matrix in the RA/Dec.\ basis, is used throughout the \textsc{fhd} analysis because it corresponds with the coherency vector defined in Equation \ref{eq:coherency_def}. However, it is more typical to report models of the polarized antenna response in terms of $\jonesmat^\text{ZA}_j(\skypos)$ because the RA/Dec.\ basis is dependent on the antenna's location on the Earth and varies with latitude.

\begin{figure*}
    \centering
    \includegraphics[width=0.7\columnwidth]{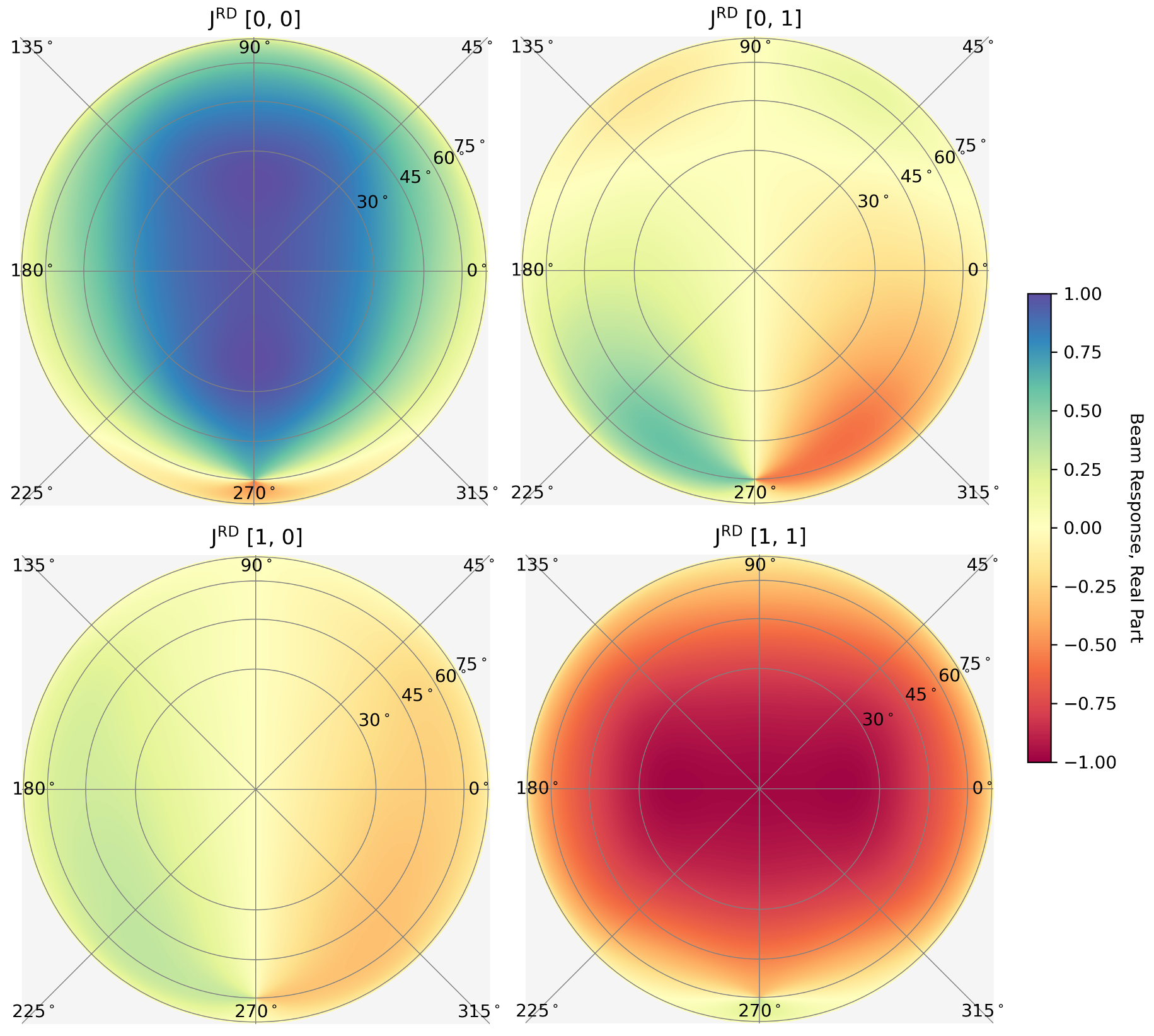}
    \caption{The Jones matrix of an MWA tile, plotted in Figure \ref{fig:jones_za}, now recast in the RA/Dec.\ coordinate system (see Equation \ref{eq:jones_rot}). Once again, the top and bottom rows correspond to the polarized responses of the east-west and north-south aligned antenna polarization, respectively. However, the left column now corresponds to the tile's response to emission polarized in the RA direction while the right column corresponds to the response to emission polarized in the Dec.\ direction. As in Figure \ref{fig:jones_za}, we plot the real part of the Jones matrix elements only. Since the RA/Dec.\ coordinate system has poles at the North and South Poles, we see a discontinuity at the bottom edge of the plots, corresponding to the position of the South Pole relative to the MWA's $-27^\circ$ latitude.}
    \label{fig:jones_ra}
\end{figure*}

In Figure \ref{fig:jones_ra} we plot the elements of the MWA Jones matrix in RA/Dec.\ coordinates by applying the transformation in Equation \ref{eq:jones_rot} to the Jones matrix presented in Figure \ref{fig:jones_za}. Once again, we plot the real part of the Jones matrix elements only. Note that we no longer see a discontinuity at zenith. Instead, the Jones matrix elements exhibit a discontinuity near the lower edge of the plots, corresponding to the position of the South Celestial Pole.

\subsection{The Mueller Matrix}

The Mueller matrix is a $4\times4$ complex matrix formed by the Kronecker product of two Jones matrices:
\begin{equation}
    \muellermat_{jk}(\skypos) = \jonesmat^{\racoord \deccoord}_j(\skypos) \otimes {\jonesmat^{\racoord \deccoord}_k}^*(\skypos)
\end{equation}
\citep{Hamaker1996a}. Here we define the Mueller matrix with respect to the RA/Dec.\ basis to once again align with the convention in Equation \ref{eq:coherency_def}. 

The Mueller matrix defines the mapping between the sky signal and visibilities. We can write the visibilities formed from correlating antennas $j$ and $k$ in terms of the Mueller matrix:
\begin{equation}
    \data_{jk} = \int d^2\skypos \,
    \muellermat_{jk}(\skypos) \coherency(\skypos) \, e^{2 \pi i \skypos \cdot \uvcoord_{jk}}.
\label{eq:visibility_definition_polarized}
\end{equation}
Here $\uvcoord_{jk}$ is the baseline vector for antennas $j$ and $k$ in units of wavelengths and $\data_{jk}$ is a 4-element vector with components
\begin{equation}
    \data_{jk} = \begin{bmatrix}
    \datascalar_{jk\ewinstpol \ewinstpol} \\
    \datascalar_{jk\nsinstpol \nsinstpol} \\
    \datascalar_{jk\ewinstpol \nsinstpol} \\
    \datascalar_{jk\nsinstpol \ewinstpol}
    \end{bmatrix}
\label{eq:vis_vector_def}
\end{equation}
where, for example, $\datascalar_{jk\ewinstpol \nsinstpol}$ represents the visibility formed by correlating the $\ewinstpol$-polarized signal from antenna $j$ with the $\nsinstpol$-polarized signal from antenna $k$, and $\ewinstpol$ and $\nsinstpol$ once again refer to the two instrumental polarizations of a dual-polarized antenna. The integral is taken across the visible sky, and the Mueller matrix $\muellermat_{jk}(\skypos)$ contains the appropriate visibility normalization. Polarized imaging, discussed in \S\ref{s:polarized_imaging}, amounts to inverting Equation \ref{eq:visibility_definition_polarized} to estimate $\coherency(\skypos)$ from the measured visiblities.

Equation \ref{eq:visibility_definition_polarized} assumes a coplanar array, where each baseline can be described by the two-element vector $\uvcoord_{jk}$. Non-coplanar baselines introduce an additional phase term in the integrand of Equation \ref{eq:visibility_definition_polarized}, but \textsc{fhd} neglects this term and assumes a reasonably coplanar array. Moderate non-coplanarity can be corrected with w-projection \citep{Cornwell1992}.

\section{The Instrumental Basis}
\label{s:instr_basis}

\textsc{fhd} reconstructs images in the ``instrumental'' polarization basis, defined as the basis that diagonalizes the Jones and Mueller matrices. Physically, we can interpret the basis vectors as the polarization directions of maximal instrumental response. It is fundamentally a non-orthogonal basis, but with good knowledge of the instrument's polarized response we can freely convert between the instrumental basis and orthogonal bases such as the RA/Dec.\ coordinate system. In this section we define the instrumental basis; \S\ref{s:polarized_imaging} explains how we use this basis for image reconstruction.

\begin{figure*}
    \centering
    \includegraphics[width=0.7\columnwidth]{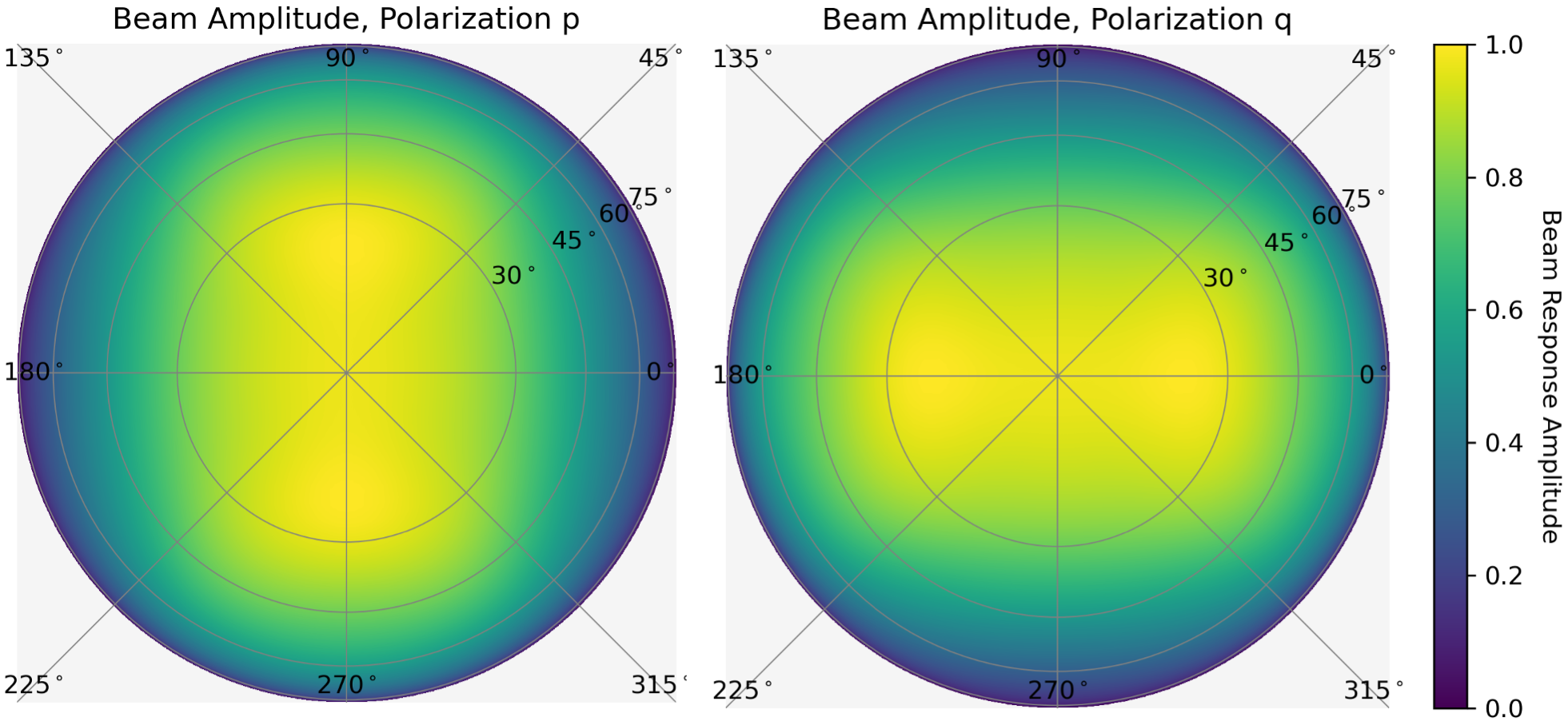}
    \caption{The sensitivity, or beam amplitude, of east-west (left) and north-south (right) aligned antenna polarizations for an MWA tile. The full Jones matrix for this tile is plotted in Figures \ref{fig:jones_za} and \ref{fig:jones_ra}. The quantities plotted here are the diagonal elements of $\antresmat_j(\skypos)$ (Equation \ref{eq:diag_response_mat}). Here they are normalized such that each response has a peak amplitude of one.}
    \label{fig:beam_amp}
\end{figure*}

The instrumental basis is defined by decomposing the Jones matrix into a product of two matrices:
\begin{equation}
    \jonesmat^{\racoord \deccoord}_j(\skypos) = \antresmat_j(\skypos) \basistransform_j(\skypos).
\label{eq:jones_decomp}
\end{equation}
$\antresmat_j(\skypos)$ is a diagonal matrix that encodes the amplitude of the instrumental response:
\begin{equation}
    \antresmat_j(\skypos) = \begin{bmatrix}
    \sqrt{|\jonesscalar_{\racoord \ewinstpol j}(\skypos)|^2 + |\jonesscalar_{\deccoord \ewinstpol j}(\skypos)|^2 } & 0 \\
    0 & \sqrt{|\jonesscalar_{\racoord \nsinstpol j}(\skypos)|^2 + |\jonesscalar_{\deccoord \nsinstpol j}(\skypos)|^2}
    \end{bmatrix}
\label{eq:diag_response_mat}
\end{equation}
where the elements of the Jones matrix are given by
\begin{equation}
    \jonesmat^{\racoord \deccoord}_j(\skypos) = \begin{bmatrix}
    \jonesscalar_{\racoord \ewinstpol j}(\skypos) & \jonesscalar_{\deccoord \ewinstpol j}(\skypos) \\
    \jonesscalar_{\racoord \nsinstpol j}(\skypos) & \jonesscalar_{\deccoord \nsinstpol j}(\skypos)
    \end{bmatrix}.
\end{equation}
The two elements of $\antresmat_j(\skypos)$ give the sensitivity of each the $\ewinstpol$ and $\nsinstpol$ polarizations of antenna $j$ to unpolarized emission from location $\skypos$ on the sky. Figure \ref{fig:beam_amp} plots these quantities for an MWA tile.

\begin{figure}
    \centering
    \includegraphics[width=\columnwidth]{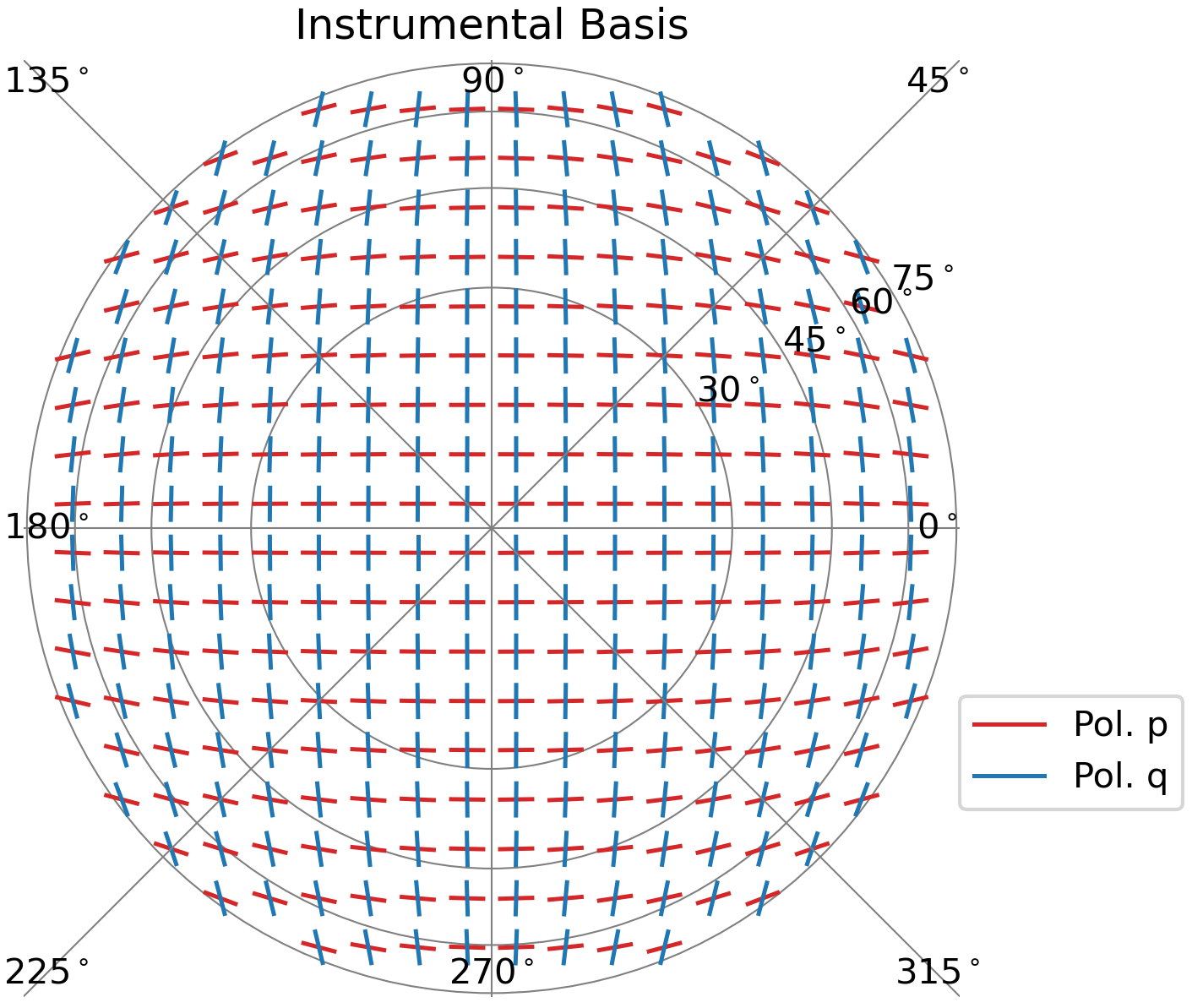}
    \caption{The instrumental basis of the MWA, as defined by the Jones matrix model plotted in Figures \ref{fig:jones_za} and \ref{fig:jones_ra}. The instrumental basis transformation is encoded in the matrix $\basistransform(\skypos)$ (see Equation \ref{eq:basis_transform}). The red line segments indicate the polarization direction that induces a maximal response in the $\ewinstpol$, or east-west aligned, antenna polarization; the blue line segments indicate the polarization direction that induces a maximal response in the $\nsinstpol$, or north-south aligned, antenna polarization. Note that the instrumental basis vectors are approximately orthogonal near zenith but are non-orthogonal off-axis.}
    \label{fig:instr_basis}
\end{figure}

While $\antresmat_j(\skypos)$ captures the instrument's response to unpolarized emission, polarized emission preferentially couples with a particular instrumental polarization. $\basistransform_j(\skypos)$ captures this effect, encoding the polarization-dependent component of the instrumental response. For a complex-valued Jones matrix, $\basistransform_j(\skypos)$ additionally encodes the complex phase.
If $\basistransform_j(\skypos)$ is identical for all antennas, then we can let $\basistransform_j(\skypos) = \basistransform(\skypos)$ for all antennas $j$. We then define a new ``instrumental basis'' on the sky, where $\basistransform(\skypos)$ transforms between the usual RA/Dec.\ basis and our new instrumental basis:
\begin{equation}
    \begin{bmatrix}
    \unitvec_\ewinstpol(\skypos) \\
    \unitvec_\nsinstpol(\skypos)
    \end{bmatrix} = \basistransform(\skypos) \begin{bmatrix}
    \unitvec_\racoord(\skypos) \\
    \unitvec_\deccoord(\skypos)
    \end{bmatrix}.
\label{eq:basis_transform}
\end{equation}
Here $\unitvec_\ewinstpol(\skypos)$ and $\unitvec_\nsinstpol(\skypos)$ are unit vectors aligned with the polarization directions that produce the maximal response from the $\ewinstpol$ and $\nsinstpol$ polarizations of each antenna. $\unitvec_{\ewinstpol}(\skypos)$ and $\unitvec_{\nsinstpol}(\skypos)$ are generally not orthogonal, so $\basistransform(\skypos)$ is not a unitary matrix. Figure \ref{fig:instr_basis} plots the instrumental basis for the MWA.

We can always define a consistent instrumental basis across a homogeneous array, where each antenna has an identical response. However, an array does not need to be homogeneous for $\basistransform_j(\skypos)$ to be invariant across antennas $j$. Antennas can have different response amplitudes across the sky provided each antenna is maximally sensitive to the same polarization directions. We cannot define an instrumental basis for arrays where antennas are rotated with respect to one another. This can also pose issues for arrays with very long baselines, where the curvature of the Earth begins to have an appreciable effect \citep{Tasse2013}.

Just as we decomposed the Jones matrix into two components in Equation \ref{eq:jones_decomp}, we represent the Mueller matrix as the product
\begin{equation}
\begin{split}
    \muellermat_{jk}(\skypos) = \beammat_{jk}(\skypos) \muellerbasistransform(\skypos)
\end{split}
\end{equation}
where
\begin{equation}
    \muellerbasistransform(\skypos) = \basistransform(\skypos) \otimes \basistransform^*(\skypos)
\end{equation}
and
\begin{equation}
    \beammat_{jk}(\skypos) = \antresmat_j(\skypos) \otimes \antresmat_k(\skypos).
\label{eq:beammat_def}
\end{equation}
Here $\beammat_{jk}(\skypos)$ is a diagonal $4 \times 4$ matrix that defines the baseline response amplitude, or beam, of baseline $\{j, k\}$. Note that Equation \ref{eq:beammat_def} does not include a complex conjugation simply because Equation \ref{eq:diag_response_mat} explicitly defines $\antresmat(\skypos)$ to be real. 

We define a new coherency vector in the instrumental basis
\begin{equation}
    \coherency_\text{inst}(\skypos) = \begin{bmatrix}
    \langle |\electricfield_\ewinstpol(\skypos)|^2 \rangle \\ \langle |\electricfield_\nsinstpol(\skypos)|^2 \rangle \\
    \langle \electricfield_\ewinstpol(\skypos) \electricfield_\nsinstpol^*(\skypos) \rangle \\
    \langle \electricfield_\ewinstpol^*(\skypos) \electricfield_\nsinstpol(\skypos) \rangle
    \end{bmatrix} = \muellerbasistransform(\skypos) \coherency(\skypos)
\label{eq:inst_coherency_def}
\end{equation}
where $\electricfield_\ewinstpol(\skypos)$ and $\electricfield_\nsinstpol(\skypos)$ are the components of the electric field aligned with unit vectors $\unitvec_\ewinstpol(\skypos)$ and $\unitvec_\nsinstpol(\skypos)$, respectively.

\textsc{fhd}'s polarized analysis pipeline consists of reconstructing the instrumental coherency $\coherency_\text{inst}(\skypos)$ and then transforming into the true coherency $\coherency(\skypos)$ by inverting Equation \ref{eq:inst_coherency_def}. We then calculate the Stokes polarization parameters using Equation \ref{eq:stokes_def}.

\section{Imaging}
\label{s:polarized_imaging}

\textsc{fhd}'s imaging pipeline follows the optimal mapmaking, or A-projection, formalism \citep{Bhatnagar2008, Morales2009}. In this approach visibilities are gridded to the \textit{uv} plane using a gridding kernel equal to the Fourier transformed beam model. \textsc{fhd} grids in the instrumental basis, forming \textit{uv} planes corresponding to each instrumental polarization mode. We pixelate the \textit{uv} plane at half wavelength spacing to enable an accurate horizon-to-horizon reconstruction of the sky signal. The gridding kernel is over-resolved with much finer pixelization to reduce spectral contamination of the reconstructed signal \citep{Barry2019a, Offringa2019, Barry2022}. Observations are processed as snapshot images (MWA observations generally have a duration of about 2 minutes), and individual snapshots are combined in the image plane \citep{Beardsley2016}.

\subsection{Gridding}

Gridding reconstructs the \textit{uv} plane from the visibilities. Here we describe gridding the $\ewinstpol \nsinstpol$-polarized visibilities; for the $\ewinstpol \ewinstpol$, $\nsinstpol \nsinstpol$, and $\nsinstpol \ewinstpol$ polarizations, simply substitute for the subscripts $\ewinstpol \nsinstpol$ below. 

For a given baseline formed by correlating signals from antennas $j$ and $k$, we denote the $\ewinstpol \nsinstpol$-polarized beam model $\beamscalar_{jk\, \ewinstpol \nsinstpol}(\skypos)$. Note that this represents one element of the $4\times4$ diagonal matrix $\beammat_{jk}(\skypos)$ presented in Equation \ref{eq:beammat_def}. Under optimal mapmaking, the gridding kernel is the Fourier transform of this beam model, which we will denote $\widetilde{\beamscalar}_{jk\, \ewinstpol \nsinstpol}(\uvcoord)$. Here $\uvcoord$ represents the \textit{uv} plane coordinate, which is Fourier dual to the sky coordinate $\skypos$ and has units of wavelengths. The tilde indicates the Fourier transformed quantity: $\widetilde{\beamscalar}_{jk\, \ewinstpol \nsinstpol}(\uvcoord) = \ft \left[ \beamscalar_{jk\, \ewinstpol \nsinstpol}(\skypos) \right]$.

Gridding with this kernel produces the following reconstructed \textit{uv} plane:
\begin{equation}
    \hat{\widetilde{\coherencyscalar}}_{\text{app}\, \ewinstpol \nsinstpol} (\uvcoord) = \frac{\sum_{jk} \widetilde{\beamscalar}_{jk\, \ewinstpol \nsinstpol}(\uvcoord_{jk}-\uvcoord) \, \datascalar_{jk\, \ewinstpol \nsinstpol}}{\sum_{jk} \widetilde{\beamscalar}_{jk\, \ewinstpol \nsinstpol }(\uvcoord_{jk}-\uvcoord)}.
\label{eq:polarized_gridding}
\end{equation}
Here $\hat{\coherencyscalar}_{\text{app}\, \ewinstpol \nsinstpol}(\skypos)$ is the apparent sky in the $\ewinstpol \nsinstpol$ polarization, defined as the estimate of the sky multiplied by the beam amplitude; $\hat{\widetilde{\coherencyscalar}}_{\text{app}\, \ewinstpol \nsinstpol} (\uvcoord)$ is its Fourier transform. The hat symbol $ \, \hat{\,} \,$ indicates that this is the reconstructed estimate. $j$ and $k$ index antennas, and $\sum_{jk}$ denotes the sum over all baselines.

The numerator in Equation \ref{eq:polarized_gridding} describes the gridding operation. If the model accurately captures the true instrumental beam, this method produces an optimal and lossless sky estimate \citep{Tegmark1997, Bhatnagar2008, Morales2009}. Errors in the beam model produce errors in the reconstructed intensities, polarization, and, in rare cases, the positions of sources. Precision beam modeling is therefore an active area of research (\citealt{Newburgh2014, Sutinjo2015, Berger2016, Wayth2016, Sokolowski2017, Line2018, Fagnoni2021a, Chokshi2021}; etc.).

The denominator in Equation \ref{eq:polarized_gridding} is known as the \textit{uv} weights and is equivalent to gridding each visibility with a value of unity. If the instrument does not have complete \textit{uv} plane measurement coverage, the weights will be equal to zero in certain \textit{uv} plane locations. We set the expression in Equation \ref{eq:polarized_gridding} equal to zero where the weights are zero. Other imaging approaches introduce regularization methods to, in effect, interpolate over incomplete \textit{uv} measurements (see \citealt{Johnson2017}, \citealt{Eastwood2018}, and \citealt{Eastwood2019}).

Equation \ref{eq:polarized_gridding} represents just one \textit{uv} weighting scheme. \textsc{fhd} supports alternative weightings as well, corresponding to different denominators in Equation \ref{eq:polarized_gridding}. For example, the denominator can be replaced with unity (often called ``natural weighting'') or with the number of visiblities that contribute to each \textit{uv} pixel (often called ``uniform weighting''). However, Equation \ref{eq:polarized_gridding} presents a sensible choice in which the reconstructed \textit{uv} pixels are scaled by the measurement sensitivity. This weighting scheme, which we call ``optimal weighting,'' was used to produce the diffuse maps in \citealt{Byrne2022}.

Because $\beammat_{jk}(\skypos)$ is defined to be diagonal, $\ewinstpol \nsinstpol$-polarized visiblities contribute to the $\ewinstpol \nsinstpol$ \textit{uv} plane only, and likewise the other visibility polarizations contribute only to their respective \textit{uv} planes. This means that each visibility is only gridded once. This is a unique feature of the instrumental basis, as any other basis choice introduces off-diagonal elements of $\beammat_{jk}(\skypos)$ and requires that each visibility is gridded four times. This holds true even when the Mueller matrix has significant off-diagonal components: the instrumental basis need not be orthogonal. Because gridding is the most computationally intensive step in the imaging pipeline, the instrumental basis allows for significant computational savings.

\subsection{Image Reconstruction}

Transforming the gridded \textit{uv} plane into sky coordinates gives
\begin{equation}
    \hat{\coherencyscalar}_{\text{app}\, \ewinstpol \nsinstpol} (\skypos) = \ft^{-1} \left[ \hat{\widetilde{\coherencyscalar}}_{\text{app}\, \ewinstpol \nsinstpol} (\uvcoord) \right],
\label{eq:app_sky_reconstruction}
\end{equation}
where $\ft^{-1}$ is the inverse Fourier transform operator.
This apparent sky map can then be converted into an estimate of the true sky, in the instrumental $\ewinstpol \nsinstpol$ polarization, by undoing the beam weighting:
\begin{equation}
    \hat{\coherencyscalar}_{\text{inst}\, \ewinstpol \nsinstpol} (\skypos) = \frac{1}{\beamscalar_{\text{avg}\, \ewinstpol \nsinstpol}(\skypos)} \ft^{-1} \left[ \hat{\widetilde{\coherencyscalar}}_{\text{app}\, \ewinstpol \nsinstpol} (\uvcoord) \right].
\end{equation}
Here $\hat{\coherencyscalar}_{\text{inst}\, \ewinstpol \nsinstpol} (\skypos)$ is an element of the instrumental coherency estimate, as defined in Equation \ref{eq:inst_coherency_def}. $\beamscalar_{\text{avg}\, \ewinstpol \nsinstpol}(\skypos) = \langle \beamscalar_{jk\, \ewinstpol \nsinstpol}(\skypos) \rangle$ is the average $\ewinstpol \nsinstpol$ beam amplitude, averaged across baselines. For a homogeneous array, $\beamscalar_{jk\, \ewinstpol \nsinstpol}(\skypos) = \beamscalar_{\text{avg}\, \ewinstpol \nsinstpol}(\skypos)$ for all baselines $\{j, k\}$.

Gridding all four visibility polarizations, $\ewinstpol \ewinstpol$, $\nsinstpol \nsinstpol$, $\ewinstpol \nsinstpol$ and $\nsinstpol \ewinstpol$ and Fourier transforming their respective \textit{uv} planes produces an estimate of the instrumental coherency $\hat{\coherency}_{\text{inst}} (\skypos)$. This is a reconstruction of the polarized sky signal defined with respect to the non-orthogonal instrumental polarization basis, plotted for the MWA in Figure \ref{fig:instr_basis}. For science applications, we convert this instrumental coherency into a fixed, orthogonal basis on the sky or Stokes parameters.

From Equation \ref{eq:inst_coherency_def}, we transform the estimated instrumental coherency into the RA/Dec.\ basis coherency:
\begin{equation}
    \hat{\coherency}(\skypos) = \muellerbasistransform^{-1}(\skypos) \hat{\coherency}_{\text{inst}} (\skypos).
\end{equation}
In general $\muellerbasistransform^{-1}(\skypos)$ is well-defined and can be calculated by numerically inverting $\muellerbasistransform(\skypos)$. However, $\muellerbasistransform(\skypos)$ is singular when the instrumental basis vectors are parallel, as is the case for the MWA at the horizon. Above the horizon $\muellerbasistransform(\skypos)$ is invertible. 

Note that the coherency estimate $\hat{\coherency}(\skypos)$ must be calculated from the instrumental coherency $\hat{\coherency}_{\text{inst}} (\skypos)$, \textit{not} from the apparent sky estimate $\hat{\coherency}_\text{app} (\skypos)$. Because the instrumental beam amplitude is polarization-dependent (see Figure \ref{fig:beam_amp}), the apparent sky estimate is intrinsically defined in the instrumental polarization basis. 

Finally, we calculate the Stokes parameters from the coherency estimate $\hat{\coherency}(\skypos)$ via Equation \ref{eq:stokes_def}:
\begin{equation}
    \begin{bmatrix}
    \hat{I}(\skypos) \\
    \hat{Q}(\skypos) \\
    \hat{U}(\skypos) \\
    \hat{V}(\skypos)
    \end{bmatrix} = \begin{bmatrix}
	1 & 1 & 0 & 0 \\
	1 & -1 & 0 & 0 \\
	0 & 0 & 1 & 1 \\
	0 & 0 & i & -i \\
	\end{bmatrix} \hat{\coherency}(\skypos).
\label{eq:stokes_reconstruction}
\end{equation}
This produces an optimal, lossless estimate of the fully-polarized true sky signal.

\begin{figure*}
    \centering
    \includegraphics[width=\columnwidth]{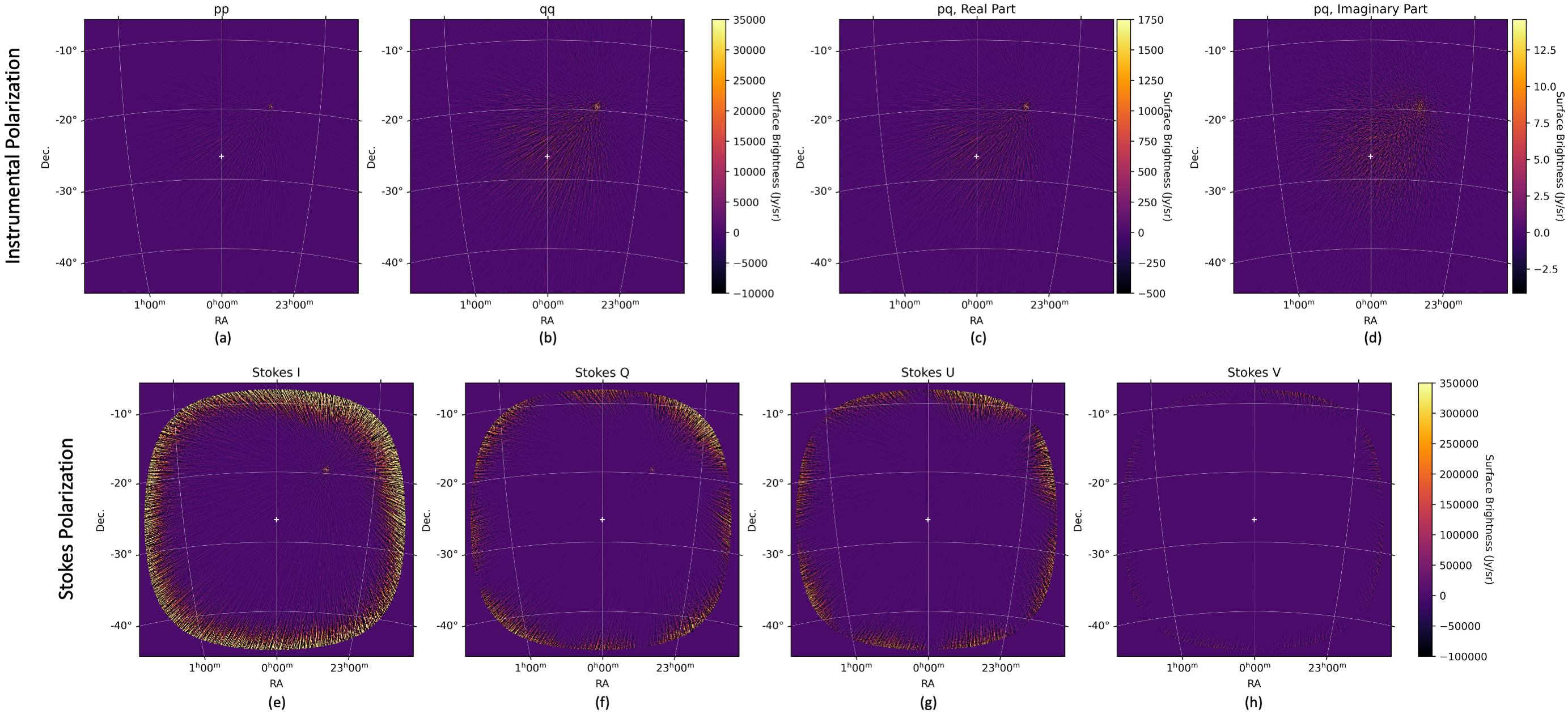}
    \caption{Example of polarized image reconstruction with \textsc{fhd}, based on a simulation of a single Stokes Q polarized point source with a polarization fraction of 50\%. The source is located in the upper right quadrant of each image at a zenith angle of 10$^\circ$ (RA 23h30m04s, Dec.\ -19.4$^\circ$). Zenith is marked with a plus symbol. Visibilities were simulated with the \textsc{pyuvsim} simulation package \citep{Lanman2019} and based on the MWA Phase I at 167-198 MHz. The visibilities were then gridded and imaged with \textsc{fhd} to produce instrumental polarization images (top row; see Equation \ref{eq:app_sky_reconstruction}) and Stokes images (bottom row; see Equation \ref{eq:stokes_reconstruction}). The instrumental polarized images $\ewinstpol \ewinstpol$ and $\nsinstpol \nsinstpol$ are real-valued, but the $\ewinstpol \nsinstpol$ and $\nsinstpol \ewinstpol$ images are complex-valued and complex conjugates of one another. We therefore plot the real and imaginary components of the $\ewinstpol \nsinstpol$-polarized image and do not plot the $\nsinstpol \ewinstpol$-polarized image. While the simulated source appears predominantly in the $\ewinstpol \ewinstpol$ and $\nsinstpol \nsinstpol$ images, a small amount of power couples into the $\ewinstpol \nsinstpol$ image as a result of non-orthogonality of the instrumental basis at the source location. The reconstructed Stokes images have a 49.52\% total polarization fraction and 49.47\% Stokes Q polarization fraction at the location of the simulated source.}
    \label{fig:image_reconstruction}
\end{figure*}

A deconvolution step can be added to the image reconstruction pipeline to reduce imaging artifacts from the array's Point Spread Function (PSF). Deconvolution assumes that sources are compact on the sky and thereby differentiates between true source emission and the PSF. \citealt{Sullivan2012} describes \textsc{fhd}'s deconvolution algorithm. Deconvolution may not be necessary or desirable when imaging large-scale, diffuse structure, or when processing data that nearly completely samples the \textit{uv} plane, as in \citealt{Byrne2022}. It should be noted that, without mitigation through deconvolution, the imaging approach described in this section produces images that are convolved with the array's PSF, as in Figure \ref{fig:image_reconstruction}.

Figure \ref{fig:image_reconstruction} presents an example of \textsc{fhd}'s image reconstruction with a simulated polarized point source. Visibilities were simulated with the \textsc{pyuvsim} simulation package\footnote{\texttt{https://github.com/RadioAstronomySoftwareGroup/pyuvsim}} \citep{Lanman2019}. Although \textsc{fhd} itself performs polarized source simulation, \textsc{pyuvsim} enables very accurate, semi-analytical interferometric simulations and serves as an independent verification of the \textsc{fhd}'s polarized imaging pipeline. The simulation is based on a zenith-pointed observation made with the MWA Phase I at 167-198 MHz. We simulated a 10 Jy point source located north-west of zenith at a zenith angle of $10^\circ$ and polarized in Stokes Q with a polarization fraction of 50\%. The simulated visibilities were gridded and imaged with \textsc{fhd}, and Figure \ref{fig:image_reconstruction} depicts the resulting images in instrumental polarization (top row) and Stokes (bottom row). The images are not deconvolved, so the simulated source appears convolved with the array's PSF.

The top row of Figure \ref{fig:image_reconstruction} presents images of the simulated source in instrumental polarization, corresponding to the quantity $\hat{\coherency}_{\text{app}} (\skypos)$ from Equation \ref{eq:app_sky_reconstruction}. Although the simulated source appears primarily in the $\ewinstpol \ewinstpol$ and $\nsinstpol \nsinstpol$ polarizations, we see a small amount of power in the $\ewinstpol \nsinstpol$-polarized image as well. The source's polarization preferentially couples power into the $\nsinstpol \nsinstpol$ image, and the $\nsinstpol \nsinstpol$ image therefore has a greater amplitude than the $\ewinstpol \ewinstpol$ image. The power in the $\ewinstpol \nsinstpol$ image is due to non-orthogonality of the instrumental polarization basis.

The bottom row of Figure 6 depicts the reconstructed Stokes images ($\hat{I}(\skypos)$, $\hat{Q}(\skypos)$, $\hat{U}(\skypos)$, and $\hat{V}(\skypos)$ from Equation \ref{eq:stokes_reconstruction}). At the source location, the reconstructed images have a polarization fraction of 49.52\%, of which 49.47\% appears in Stokes Q. This aligns well with the simulation input of 50\% fractional polarization in Stokes Q.

\subsection{Comparison with Other Widefield Polarized Imagers}

\textsc{fhd}'s polarized imaging pipeline joins the field of other polarized widefield imagers including \textsc{wsclean} \citep{Offringa2014} and the \textsc{rts} \citep{Mitchell2008}. It is perhaps most similar to the imaging approach discussed in \citealt{Tasse2013}, which, like \textsc{fhd}, is based on optimal mapmaking (A-projection) and grids visibilities with a kernel derived from the beam model.

\citealt{Tasse2013} describes an analysis pipeline built for widefield imaging with LOFAR. Although LOFAR's station configurations differ across the array, each station has identical polarization alignment. This enables the analysis to define an implicit instrumental basis for visibility gridding, such that each visibility is gridded to just one \textit{uv} plane. \citealt{Tasse2013} transforms between the the instrumental basis and an orthogonal polarization basis in the \textit{uv} plane, applying a correction factor to the gridded visibilities before deconvolving. This is in contrast to \textsc{fhd}, which performs this transformation in the image plane after dividing by the instrumental beam amplitude.

\textsc{wsclean} \citep{Offringa2014} is a fully-polarized widefield imager that was originally built for the MWA but has been widely utilized for analysis of data from instruments including LOFAR, GMRT, the Very Large Array (VLA), the OVRO-LWA, and the Australian Square Kilometre Array Pathfinder (ASKAP). It was not developed as an A-projection algorithm, although it supports a variety of gridding kernels and can grid with the instrumental beam model \citep{Lynch2021}. In its polarized imaging mode, \textsc{wsclean} produces instrumentally polarized $\ewinstpol \ewinstpol$, $\nsinstpol \nsinstpol$, $\ewinstpol \nsinstpol$ and $\nsinstpol \ewinstpol$ images. Much like in \textsc{fhd}, it then follows with a beam correction step that transforms the images into Stokes parameters. This step corrects for direction-dependent image attenuation from the beam amplitude and gridding kernel and performs the polarization basis transformation.

\citealt{Mitchell2008} describes the \textsc{rts}, a GPU-accelerated calibration and imaging pipeline for the MWA, also discussed in \citealt{Tingay2013}. This pipeline also performs widefield imaging and produces fully-polarized images. Much like \textsc{wsclean} and \textsc{fhd}, it produces images in the instrumental polarization basis and then uses the Mueller matrix to transform to Stokes parameters. Like \textsc{wsclean}, it was not initially developed to perform A-projection, and it typically does not use the instrumental beam for gridding.

\section{Polarized Calibration}
\label{s:polarized_calibration}

\textsc{fhd} performs direction-independent sky-based calibration. This calibration approach is based on the measurement equation which, in its fully polarized form, is given by
\begin{equation}
    \data_{jk} = \gainsmat_{jk} \modelvals_{jk} + \boldsymbol{n}_{jk}
\end{equation}
\citep{Hamaker1996a}. Here $\data_{jk}$ is the 4-element measured visibility vector given by Equation \ref{eq:vis_vector_def}, $\modelvals_{jk}$ is the model visiblities derived from a model of the sky and simulated through a model of the instrument response, and $\boldsymbol{n}_{jk}$ is the noise on the measurement. $\gainsmat_{jk}$ is a $4\times4$ gain matrix. The gain matrix serves as a direction-independent modification of the Mueller matrix. Empirically fitting the gain terms constrains uncertainties in the modeled Mueller matrix, helping it to better align with the true Mueller matrix that governs the instrumental response. The measurement equation is implicitly defined per time step and frequency channel.

\subsection{Gain Parameterization}

In its most general form, $\gainsmat_{jk}$ is given by
\begin{equation}
    \gainsmat_{jk} = \begin{bmatrix}
    \gain_{j\ewinstpol\ewinstpol} \, \gain_{k\ewinstpol\ewinstpol}^* & \gain_{j\ewinstpol\nsinstpol} \, \gain^*_{k\ewinstpol\nsinstpol} & \gain_{j\ewinstpol\ewinstpol} \, \gain^*_{k\ewinstpol\nsinstpol} & \gain_{j\ewinstpol\nsinstpol} \, \gain^*_{k\ewinstpol\ewinstpol} \\
    \gain_{j\nsinstpol\ewinstpol} \, \gain^*_{k\nsinstpol\ewinstpol} & \gain_{j\nsinstpol\nsinstpol} \, \gain^*_{k\nsinstpol\nsinstpol} & \gain_{j\nsinstpol\ewinstpol} \, \gain^*_{k\nsinstpol\nsinstpol} & \gain_{j\nsinstpol\nsinstpol} \, \gain^*_{k\nsinstpol\ewinstpol} \\
    \gain_{j\ewinstpol\ewinstpol} \, \gain^*_{k\nsinstpol\ewinstpol} & \gain_{j\ewinstpol\nsinstpol} \, \gain^*_{k\nsinstpol\nsinstpol} & \gain_{j\ewinstpol\ewinstpol} \, \gain^*_{k\nsinstpol\nsinstpol} & \gain_{j\ewinstpol\nsinstpol} \, \gain^*_{k\nsinstpol\ewinstpol} \\
    \gain_{j\nsinstpol\ewinstpol} \, \gain^*_{k\ewinstpol\ewinstpol} & \gain_{j\nsinstpol\nsinstpol} \, \gain^*_{k\ewinstpol\nsinstpol} & \gain_{j\nsinstpol\ewinstpol} \, \gain^*_{k\ewinstpol\nsinstpol} & \gain_{j\nsinstpol\nsinstpol} \, \gain^*_{k\ewinstpol\ewinstpol}
    \end{bmatrix}.
\end{equation}
Here $\gain_{j\ewinstpol\ewinstpol}$ and $\gain_{j\nsinstpol\nsinstpol}$ are the complex gains of the $\ewinstpol$ and $\nsinstpol$ polarizations of antenna $j$, respectively. $\gain_{j\ewinstpol\nsinstpol}$ and $\gain_{j\nsinstpol\ewinstpol}$ are the cross gains. $\gain_{j\ewinstpol\nsinstpol}$, for example, denotes the degree to which signal we expect to appear only in the $\nsinstpol$ polarization of antenna $j$ also appears in the $\ewinstpol$ polarization \citep{Sault1996}.

The calibration solutions are then calculated by minimizing a cost function given by
\begin{equation}
    \chi^2(\gains) = \sum_{jk} \left( \data_{jk} - \gainsmat_{jk} \modelvals_{jk} \right)^\dag \left( \data_{jk} - \gainsmat_{jk} \modelvals_{jk} \right),
\label{eq:polarized_cal_with_gain_crosses}
\end{equation}
where $\sum_{jk}$ denotes the sum over all baselines.
Here we omit all frequency dependence and assume that the cost function is minimized independently for each frequency channel. For discussion of \textsc{fhd}-based precision bandpass calibration techniques, see \citealt{Barry2019a, Barry2019b, Li2019}.

If an instrument experiences minimal cross-polarization signal coupling, as is the case for the MWA, we can safely set all the cross gains to zero. $\gainsmat_{jk}$ is then diagonal, and we can denote each antenna gain with a single polarization index: $\gain_{j\ewinstpol}$ and $\gain_{j\nsinstpol}$. The calibration cost function then becomes
\begin{equation}
    \chi^2(\gains) = \sum_{jk} \sum_{ab} \left| \datascalar_{jk\,ab} - \gain_{ja} \gain^*_{kb} \modelval_{jk\,ab} \right|^2,
\label{eq:polarized_cal_all_vis}
\end{equation}
where $a$ and $b$ each index the two instrumental polarization modes $\{\ewinstpol, \nsinstpol\}$. \textsc{fhd}'s polarized calibration currently does not support nonzero cross gains.

\subsection{Constraining the Cross Polarization Phase}

\textsc{fhd} initially did not use the cross polarization visibilities $\data_{\ewinstpol \nsinstpol}$ and $\data_{\nsinstpol \ewinstpol}$ in calibration. Excluding the cross polarization visibilities makes calibration separable in polarization. Two independent cost functions are minimized, corresponding to the two instrumental polarizations:
\begin{equation}
    \chi_\ewinstpol^2(\gains_\ewinstpol) = \sum_{jk} \left| \datascalar_{jk\,\ewinstpol \ewinstpol} - \gain_{j\ewinstpol} \gain^*_{k\ewinstpol} \modelval_{jk\,\ewinstpol \ewinstpol} \right|^2
\label{eq:per_pol_cal_ew}
\end{equation}
and
\begin{equation}
    \chi_\nsinstpol^2(\gains_\nsinstpol) = \sum_{jk} \left| \datascalar_{jk\,\nsinstpol \nsinstpol} - \gain_{j\nsinstpol} \gain^*_{k\nsinstpol} \modelval_{jk\,\nsinstpol \nsinstpol} \right|^2.
\label{eq:per_pol_cal_ns}
\end{equation}
However, calibrating in this way introduces a new calibration degeneracy, corresponding to the overall phase difference between the $\ewinstpol$ and $\nsinstpol$ gains across all antennas. We can identify this degeneracy by noting that the transformation $\gains_\ewinstpol \rightarrow \gains_\ewinstpol \, e^{-i\crosspolphase/2}$ and $\gains_\nsinstpol \rightarrow \gains_\nsinstpol \, e^{i\crosspolphase/2}$ does not affect the calibration solutions. We call the parameter $\crosspolphase$ the ``cross polarization phase.''

The cross polarization phase is not degenerate when we calibrate with the cross polarization visibilities, as in Equation \ref{eq:polarized_cal_with_gain_crosses} or \ref{eq:polarized_cal_all_vis} --- provided these cross visibilities are nonzero. Although it is sometimes asserted that this phase can only be constrained by calibrating to a polarized source \citep{Sault1996, Bernardi2013, Lenc2017}, in the widefield limit the nonorthogonality of the instrumental polarization basis couples appreciable unpolarized source power into the cross polarization visibilities. This allows for constraint of the cross polarization phase even while using a fully unpolarized sky model. (Note that this technique was independently developed in \citealt{Anderson_thesis}.)

When adapting \textsc{fhd} to perform fully polarized calibration, we chose to largely retain the original calibration pipeline and to supplement it with an additional step to constrain the cross polarization phase. As a result, all calibration parameters other than the cross polarization phase are fit from the single polarization visibilities $\data_{\ewinstpol \ewinstpol}$ and $\data_{\nsinstpol \nsinstpol}$. Since the original calibration pipeline constrains the relative phase of the gains across frequency \citep{Beardsley2016, Barry2019a, Barry2019b, Li2019}, the degenerate cross polarization phase amounts to a single parameter across all antennas and frequencies. If each frequency channel were calibrated independently, we would need to calculate a cross polarization phase at each frequency. We fit the cross polarization phase by plugging the solutions into the fully polarized cost function given by Equation \ref{eq:polarized_cal_all_vis}. We let $\gains_\ewinstpol = \hat{\gains}_\ewinstpol \, e^{-i\crosspolphase/2}$ and $\gains_\nsinstpol = \hat{\gains}_\nsinstpol \, e^{i\crosspolphase/2}$, where $\hat{\gains}$ are the gains calibrated up the the cross polarization phase. This gives
\begin{equation}
\begin{split}
    \chi^2(\crosspolphase) = \sum_{jk} & \left[ 
    \left| \datascalar_{jk \, \ewinstpol \nsinstpol} - e^{-i\crosspolphase} \hat{\gain}_{j \ewinstpol} \, \hat{\gain}^*_{k \nsinstpol} \, \modelval_{jk \, \ewinstpol \nsinstpol} \right|^2 \right. \\
    & \left. + \left| \datascalar_{jk \, \nsinstpol \ewinstpol} - e^{i\crosspolphase} \hat{\gain}_{j \nsinstpol} \, \hat{\gain}^*_{k \ewinstpol} \, \modelval_{jk \, \nsinstpol \ewinstpol} \right|^2
    \right].
\end{split}
\end{equation}

We can calculate the cross polarization phase analytically by finding the value $\hat{\crosspolphase}$ that minimizes $\chi^2(\crosspolphase)$. We find that
\begin{equation}
    \hat{\crosspolphase} = \argument \left[ \sum_{jk} \left( \datascalar^*_{jk \, \ewinstpol \nsinstpol} \, \hat{\gain}_{j \ewinstpol} \, \hat{\gain}^*_{k \nsinstpol} \, \modelval_{jk \, \ewinstpol \nsinstpol} + \datascalar_{jk \, \nsinstpol \ewinstpol} \, \hat{\gain}^*_{j \nsinstpol} \, \hat{\gain}_{k \ewinstpol} \, \modelval^*_{jk \, \nsinstpol \ewinstpol} \right) \right],
\end{equation}
where $\argument$ denotes the complex phase.

\section{Conclusion}

We describe an efficient and robust widefield polarized calibration and imaging pipeline implemented in the \textsc{fhd} software package. The pipeline employs an analysis approach that reconstructs images in the instrumental polarization basis. Provided all antennas have the same polarization alignment and are not rotated with respect to one another, this enables computationally efficient processing as each visibility is gridded to just one \textit{uv} plane. The pipeline implements fully-polarized calibration by supplementing \textsc{fhd}'s original calibration implementation with a constraint on the cross polarization phase.

This analysis approach accounts for all widefield polarization effects and accurately reconstructs horizon-to-horizon images---if the instrument's polarized beam is well-modeled. Imaging is susceptible to errors if the instrumental response amplitude is incorrectly modeled. Furthermore, the analysis must accurately define the instrumental polarization basis with respect to the sky coordinates. Low-level errors in the polarized beam model produces errors in the reconstructed polarization directions, which in turn leads to polarization mode-mixing. As fractional beam modeling errors can be quite large at low elevation angles, this effect can produce significant image reconstruction errors \citep{Bernardi2013, Lenc2017}. Recent analyses of MWA data have found evidence of Stokes I to Q polarization leakage of up to 40\% near the horizon due to beam modeling errors \citep{Lenc2017, Riseley2018, Byrne2022}. While polarization mode mixing can be estimated and mitigated in the image plane \citep{Lenc2016, Lenc2017, Byrne2022}, analyses would benefit from improved \textit{a priori} beam modeling.

Further extensions to \textsc{fhd}'s polarized imaging pipeline could add support for arrays with variable antenna polarization alignments. This would preclude the use of the instrumental basis for visibility gridding, increasing the computational cost of processing. However, it would extend this polarized imaging technique to a wider class of arrays, including arrays with very long baselines in which the curvature of the earth has an appreciable effect on the polarization alignment.

As noted in \S\ref{s:polarized_calibration}, we can leverage widefield projection effects to constrain polarized calibration solutions even with a fully unpolarized sky model. However, constraint of the cross polarization phase $\crosspolphase$ could be improved with calibration to a known polarized source, as in \citealt{Bernardi2013}. In addition, further investigation could explore whether calibration would benefit from using the cross polarization visibilities $\data_{\ewinstpol \nsinstpol}$ and $\data_{\nsinstpol \ewinstpol}$ to fit all calibration parameters, not just $\crosspolphase$.

The \textsc{fhd} polarized imaging pipeline builds upon the success of optimal mapmaking and A-projection imaging algorithms to enable accurate and efficient widefield polarized imaging for low-frequency radio arrays. The pipeline has produced new polarized diffuse maps with data from the MWA, presented in \citealt{Byrne2022}. Coupled with a continued investment in precision polarized beam modeling, \textsc{fhd}'s polarized imaging capabilities could pave the way for future low-frequency polarimetric studies.

\section*{Acknowledgements}

We would like to thank James Aguirre, Zachary Martinot, and Daniel Mitchell for conversations that contributed to this work. Thank you to Yuping Huang for assisting with revision of this paper. This work was directly supported by National Science Foundation (NSF) grant 1613855.


\bibliography{example}

\end{document}